\documentclass[preprint2]{aastex} 

\def \etal{{et al.}}
\def \eg{{e.g.,}}

\def \TRACE{{\it TRACE}}
\def \SOHO{{\it SOHO}}
\def \Yohkoh{{\it Yohkoh}}

\shorttitle{ Resonance Scattering }
\shortauthors{Brickhouse and Schmelz}

\begin{document}

\title{The Transparency of Solar Coronal Active Regions}

\author{N. S. Brickhouse\altaffilmark{1} and J. T. Schmelz\altaffilmark{2}}

\altaffiltext{1}{Harvard-Smithsonian Center for Astrophysics, 60 Garden St., Cambridge, MA 02138; nbrickhouse@cfa.harvard.edu}

\altaffiltext{2}{Physics Department, University of Memphis, Memphis, TN 38152}

\begin{abstract}
Resonance scattering has often been invoked to explain the
disagreement between the observed and predicted line ratios of
\ion{Fe}{17} $\lambda$15.01 to \ion{Fe}{17} $\lambda$15.26 (the
``3C/3D'' ratio). In this process photons of $\lambda$15.01, with its
much higher oscillator strength, are preferentially scattered out of
the line of sight, thus reducing the observed line ratio.  Recent
laboratory measurements, however, have found
significant inner-shell \ion{Fe}{16} lines at 15.21 and 15.26~\AA,
suggesting that the observed 3C/3D ratio results from blending.
Given our new understanding of the
fundamental spectroscopy, we have re-examined the original solar
spectra, identifying the \ion{Fe}{16} $\lambda$15.21 line and
measuring its flux to account for the contribution of \ion{Fe}{16} to
the $\lambda$15.26 flux. Deblending brings the 3C/3D ratio into good
agreement with the experimental ratio; hence, we find no need to
invoke resonance scattering.  Low  opacity in \ion{Fe}{17}
$\lambda$15.01 also implies low opacity for
Fe~XV $\lambda$284.2, ruling out resonance scattering as the cause of
the fuzziness  of \TRACE\ and \SOHO -EIT 284~\AA\
images. The images must, instead, be unresolved due to the
large number of structures at this temperature.  Insignificant
resonance scattering implies that future instruments with higher
spatial resolution could resolve the active region plasma into its
component loop structures.
\end{abstract}


\section{Introduction}

The validity of the optically thin approximation for coronal plasma
has been discussed in the literature since the beginnings of solar
X-ray and EUV spectroscopy (\eg\ Pottasch 1963, and references
therein). Resonance scattering of \ion{Fe}{17} $\lambda$15.01 ($2p^6\
^1S_0\ -\ 2p^5 3d\ ^1P_1$, known as ``3C''), in particular, has been
the subject of a long-standing controversy. Observations of the ratio
of the 3C flux to that of \ion{Fe}{17} $\lambda$15.26 ($2p^6\ ^1S_0\ -\
2p^53d\ ^3D_1$, known as ``3D'') gave ratios in the range 1.6 to 2.3
(Rugge \& McKenzie 1985; Schmelz \etal\ 1997; Saba \etal\ 1999),
whereas collisional radiative models predicted a ratio of about four
(Smith \etal\ 1985; Loulergue \& Nussbaumer 1973; Bhatia \& Doschek
1992; Cornille \etal\ 1994). Schmelz \etal\ and Saba \etal\ also
found, using the Flat Crystal Spectrometer (FCS) on the {\it Solar Maximum
Mission}, that the lowest values were preferentially at the solar
limb.  This center-to-limb effect bolstered the argument for
resonance scattering of $\lambda$15.01, given that the limb photons traverse a longer path ({\it cf,}
Phillips \etal\ 1996).

Significant new results have recently come to light: laboratory
measurements from the electron beam ion trap (EBIT) at Lawrence
Livermore National Laboratory found the 3C/3D flux ratio to be 3.04
$\pm$ 0.12 (Brown et al. 1998).  Using the National Institute of
Standards and Technology EBIT, Laming et al. (2000) confirmed the
value to be close to three. Individual measurements at different beam
energies from the two groups span a range from 2.50 to 3.15.  These
measurements imply that the amount of resonance scattering was
overestimated in the solar analyses described above.  Furthermore, new
theoretical models are converging toward a ratio closer to three
(e.g. Doron \& Behar 2002; Chen \& Pradhan 2002; Gu 2003), though all
published models continue to exceed the measurements by at least
10\%. New Dirac R-matrix calculation show excellent agreement with the
EBIT measurements (Chen 2005; Chen \& Pradhan 2005).

Brown \etal\ (2001) have also reported on experiments in which a
steady stream of neutral iron was injected into the EBIT, producing an
underionized plasma with both \ion{Fe}{16} and \ion{Fe}{17}. They
found 3C/3D flux ratios as low as 1.9 $\pm$ 0.11, and argued that
contamination of \ion{Fe}{17} $\lambda$15.26 by the inner-shell
\ion{Fe}{16} line ($2p\ (3s3d\ ^3D)\ ^2P_{3/2} -\ 3s\ ^2S_{1/2}$)
could account for the discrepancy between the laboratory ratio for a
pure Fe XVII plasma and the solar spectra. Blending as an explanation
for the solar results presumably implies that the center-to-limb
effect in the solar data is due to chance.

New observations of stellar coronae have not confirmed the solar line
ratios. Spectra for many stars observed with {\it Chandra} and {\it
XMM-Newton} yield a 3C/3D ratio of about three (Ness et al. 2003;
Audard et al. 2004; G\"{u}del et al. 2004), suggesting that stellar
coronae do {\it not} generally show resonance scattering (but see
Testa \etal\ 2004 and Matranga \etal\ 2005). Full-star observations cannot rule out resonance
scattering in individual active regions, however, since the number of
photons scattered out of the line of sight could be offset by a
similar number of photons scattered into the line of
sight. Furthermore, the sample of stars does not include stars with
coronae in the solar coronal temperature range (2--4 MK), such as
$\alpha$ Centauri or Procyon, for which blending with \ion{Fe}{16} might be
expected.

These new experimental, theoretical, and observational results
motivate us to re-investigate solar observations from the Flat Crystal
Spectrometer on the {\it Solar Maximum Mission.} The reanalysis of
these data, using new atomic data, is presented in \S 2. In \S 3 we
consider the implications of our results.  In particular, we find
that resonance scattering is not responsible for the fuzziness seen in
the solar images obtained with the 284~\AA\ passband of the {\it
Transition Region and Coronal Explorer} (\TRACE) and the EUV Imaging
Telescope (EIT) on the {\it Solar and Heliospheric Observatory}
(\SOHO).

\section{Analysis}

The data analyzed here were obtained with the FCS (Acton \etal\ 1980)
and are discussed in detail in earlier papers (Schmelz \etal\ 1997;
Saba \etal\ 1999). The instrument had a 15 arcsec field-of-view and
could scan the soft X-ray resonance lines of prominent ions in the
range of 1.5~\AA\ to 20.0~\AA\ with a spectral resolution of
0.015~\AA\ at 15 \AA. In this Letter, we reanalyze the lines from 31
of the 33 spectral scans from quiescent active regions (\ion{Fe}{17}I
$\lambda$14.21 could not be measured in two of these spectra -- see
below).  Figure 1 shows the portion of a typical FCS spectrum
containing the lines of interest. Spectra where plasma conditions were
changing significantly with time were excluded from the sample.

The top panel of Figure~2 shows the observed 3C/3D line ratio as a
function of temperature. The flux ratio of \ion{Fe}{17}I $\lambda$14.21 to
\ion{Fe}{17} $\lambda$16.78 provides a good temperature diagnostic, with
its high signal-to-noise ratio and abundance-insensitivity.
Calculations using the Astrophysical Plasma Emission Code (APEC)
version 1.3 (Smith et al. 2001) give the temperatures for each of the
individual measured flux ratios. The APEC emissivities incorporate the
ionization balance models of Mazzotta et al. (1998).  Models for the
other strong \ion{Fe}{17} lines, $\lambda$17.05 and $\lambda$17.10, are
less certain than for $\lambda$16.78, since these lines have a larger
contribution from dielectronic recombination and resulting cascades
(see Gu 2003), and hence are more dependent on the ionization state
model. The \ion{Mg}{11} and \ion{Ne}{9} G-ratios (i.e. the ratios of the sum of the
forbidden plus intercombination line fluxes to the flux of the
resonance line) are also temperature-dependent, but are of lower
signal-to-noise ratio due to the weakness of the intercombination
lines.

Most of the observed 3C/3D ratios are clustered, with significantly
less than the average laboratory value of 2.9; however, two of the
three highest temperature measurements also give higher 3C/3D ratios,
within 1$\sigma$ of 2.9. Moreover, the best-fit line to these data
shows a strong temperature-dependence, inconsistent with calculations
(Gu 2003), while the flux ratio of 3C to $\lambda$16.78
(middle panel) shows only modest temperature-dependence as expected
from calculations.  These results strongly suggest blending of $\lambda$15.26.

In the underionized EBIT plasma, Brown et al. (2001)  also measure a second
inner-shell \ion{Fe}{16} line at 15.21~\AA\ ($2p\ (3s3d^3D)\ ^2P_{1/2} -\
3s\ ^2S_{1/2}$).  We identify this line for the first time in an
active region spectrum, and use it to estimate the strength of the
\ion{Fe}{16} blend at 15.26~\AA. \ion{Fe}{16} calculations from the Hebrew
University Lawrence Livermore Atomic Code (HULLAC), which will be
available in the next APEC release, give a scaling factor of 0.83
(D. Liedahl, private communication). We note that this scaling factor
is significantly higher than the factor of 0.5 recommended by Brown et
al. (2001) based on multiconfigurational Dirac-Fock calculations, primarily due to a large
difference in the branching ratio between radiative decay and
autoionization for $\lambda$15.21~\AA.

The bottom panel of Figure 2 shows the line ratios that result after
subtracting the \ion{Fe}{16} $\lambda$15.26 blend, using \ion{Fe}{16}
$\lambda$15.21 as a proxy to determine its flux. The weighted mean of
this distribution is 2.76 $\pm$ 0.23, statistically indistinguishable
from the laboratory ratio. While the
$\lambda$15.21 proxy is, to our knowledge, unblended at temperatures
below 5~$\times 10^6$~K, its use for hotter plasmas is complicated by
the presence of \ion{Fe}{19} $\lambda$15.20 (e.g. the {\it Chandra} spectrum
of Capella, Desai et al. 2005).  It seems likely that the line
tentatively identified in the flare spectrum by Phillips et al. (1982)
as \ion{Fe}{16} may be dominated by \ion{Fe}{19} as well.

In light of this new analysis, we reconsider other examples of
full-Sun X-ray spectra which include \ion{Fe}{17} $\lambda$15.01 and
$\lambda$15.26. In particular, we expect active region
measurements to show lower line ratios and flare spectra to
approach the laboratory value.  Indeed active region measurements show
lower 3C/3D values (Blake \etal\ 1965; Evans \& Pounds 1968; Walker
\etal\ 1974), whereas flare spectra (e.g. Neupert \etal\
1973) give higher values. These results are thus consistent with
\ion{Fe}{16} blending at 15.26~\AA\ in the low-temperature active region
spectra.  They also suggest that a reduced ratio might be observed
also in high signal-to-noise ratio, high resolution spectra of stars
with cooler coronae.

\section{Discussion}

Resonance scattering has been suggested as one of the reasons for the
fuzzy appearance of the images obtained from the \ion{Fe}{15} 284~\AA\
passband of \TRACE\ and \SOHO -EIT, illustrated in
Figure~3. In this section, we discuss possible
explanations of the 284-\AA fuzziness and estimate the upper limit
for the resonance scattering contribution to this fuzziness.

A common explanation for the fuzzy appearance of the \TRACE\ and
EIT 284~\AA\ images is instrument scattering; however, measurements of
the 284~\AA\ point spread function are identical to those of the 171- and 
195~\AA\ passbands, which show much cleaner  images
(Golub et al. 1998). Furthermore, prominences seen in absorption 
have sharp edges in the 284~\AA\ images (Fig.~3, lower panel), demonstrating 
that the instrument can resolve fine structures.

A second candidate explanation is contamination of
\ion{He}{2} $\lambda$304 photons in the passband. This contamination
certainly exists
for coronal holes, which do not appear dark (as they do, for example,
in the thin Aluminum images from the Soft X-ray Telescope on
\Yohkoh); however, there should be no significant $\lambda$304-contribution in active
region areas because of the higher temperatures, or above the solar limb
where the scale height of $\lambda$304 is too small.

Resonance scattering has also been suggested as
the cause of the \TRACE\ 284~\AA\ fuzziness. The \ion{Fe}{17}
$\lambda$15.01 result allows us to estimate the contribution of
resonance scattering to \ion{Fe}{15} $\lambda$284.2. Using Chen's
(2005) calculated 3C/3D ratio of 2.85 at $log T =$ 6.4, and the
observed FCS ratio of 2.76 $\pm$ 0.21 gives an escape
probability of 2.76/2.85, corresponding to $\tau_{15.01} \approx\
-ln(0.97) =\ 0.032$. Following Acton (1978), $\tau\ \propto\ f\
A_i(T)\ \lambda $, where $f$ is the oscillator strength, $A_i$ is the
ionization fraction (which is a function of temperature), and
$\lambda$ is the wavelength. Solving for $\tau_{284.2}$, with the ionization balance at $log T =$ 6.4, gives

\begin{equation}
\tau_{284.2} \  =\ \tau_{15.01} \times {{(f_{284.2})\ (A_{\rm FeXV})\ {(284.2)}} \over {(f_{15.01})\ (A_{\rm FeXVII})\ {(15.01)}}} 
\end{equation}

\begin{equation}
\tau_{284.2} \ =\ {0.032}\ \times {{( 0.82)\  (0.188)\ (284.2)} \over { (2.73)\ (0.426)\  (15.01)}} \ =\ 0.080
\end{equation}

\noindent
This corresponds to an escape probability of 92\%, which indicates
that very few photons are available to contribute to the 284~\AA-image
fuzziness. For possible uncertainties of 10 to 15 \% in
experimental and theoretical 3C/3D ratios,
the escape probability ranges from 0.84 to 1.0, for $\tau_{15.01} < 0.17$. Using the same argument, with the relative coronal
abundances of Fludra \& Schmelz (1999),
we find that the \ion{O}{8} $\lambda$18.97 opacity is
comparable to that of $\lambda$15.01, while the opacities of \ion{Ne}{9}
$\lambda$13.46 and \ion{Mg}{11} $\lambda$9.17 are 2 to 3 times lower, confirming the
findings of Schmelz et al. (1997) that these important diagnostics are
not contaminated for use in emission measure distribution analysis.

The assumed temperature $T$, estimated from the 
measured ratios of \ion{Fe}{18} $\lambda$14.21 to \ion{Fe}{17}
$\lambda$16.78, can be further constrained by the laboratory measurements.
The experimental 3C/3D ratio was
measured as a function of the abundance ratio of \ion{Fe}{16} to \ion{Fe}{17},
providing a good fit to the theoretical predictions.  The experimental
plasma is not in collisional ionization equilibrium, due to the
continuous ionization from neutral iron up to the charge states of
interest, and moreover, the collisional processes are excited by a
narrow beam rather than a Maxwellian distribution as assumed for the
solar coronal plasma. Thus, an interpretation of the experimental line
ratio in terms of an ionization equilibrium temperature is not
strictly valid, since the resonance excitation contributions will not
be identical. Nevertheless, this estimate can give a consistency check
on the \ion{Fe}{15} population. For the laboratory value where the
populations of \ion{Fe}{16} and \ion{Fe}{17} are equal, the 3C/3D
ratio is 1.90 $\pm$ 0.11 (Brown et al. 2001), slightly below the
average of the cluster.  Thus the diagnostics all appear consistent
with each other in the blending scenario.

In collisional ionization equilibrium, such a ratio corresponds to
$log T \sim $ 6.32. For this lower temperature, $A_{\rm FeXV}/A_{\rm FeXVII}$
is about 3 times larger; however, Chen's (2005) predicted line ratio
is about 5\% less, such that our average value puts $\tau_{15.01}$
essentially at zero, with an upper limit below  0.2.
Meanwhile, loop differential emission measure (DEM) distributions tend
to be fairly flat-topped between $log T =$ 6.3 and 6.4 (e.g. Schmelz
et al. 2001), such that an intermediate value seems most reliable for
estimating the $A_{\rm FeXV}/A_{\rm FeXVII}$ ratio. We then find $\tau_{284.2}
= $ 0.12, with a plausible range from 0.0 to $<$ 0.2. For opacities in
this range, the model images of Wood \& Raymond (2000) still retain the
appearance of loops with sharp rather than blurred boundaries.

We refer to the last candidate explanation for the 284~\AA-fuzziness
as the ``filling-factor'' model, which is related to the observational result 
that coronal structures appear fuzzy if they are not resolved by the instrument
(see Fig. 3 of Deluca et al. 2005).
It is well established that the DEM of active
region and quiet Sun plasma peaks between 2 and 3 MK. In other words,
more higher temperature (284~\AA) plasma exists along the line of
sight than cooler (171~\AA) plasma, such that
the piling up of structures along the line of sight
may contribute to the fuzziness factor.
In the 171-\AA\ band ($\sim$1~MK), the 0.5-arcsec resolution of \TRACE\ revealed 
substantial substructure that had never been seen before; in the
284-\AA\ band ($\sim$2--3~MK), the spatial resolution
does not appear to be quite sufficient. 
An additional factor of two improvement in spatial resolution may be 
adequate, but 0.1-arcsec pixel size is technically feasible and seems preferable. 

We conclude, therefore, that the filling-factor model provides the most
likely explanation for the 284-image fuzziness, and thus
future instruments with higher spatial resolution may be able to resolve the
active region plasma into its component structures. Efforts to
understand the coronal heating process should benefit from resolved
images closer to the temperature at which the dominant heating occurs.

We would like to thank J. Saba, K. Nasraoui, D. Liedahl, L. Golub, and
J. Cirtain. Solar physics research at the University of Memphis is supported by NSF
ATM-0402729 and NASA NNG05GE68G. N. B. was supported by NASA 
contract NAS8-39073 to the Smithsonian Astrophysical Observatory for the
Chandra X-ray Center.

{}

\clearpage
\begin{figure}
\plotone{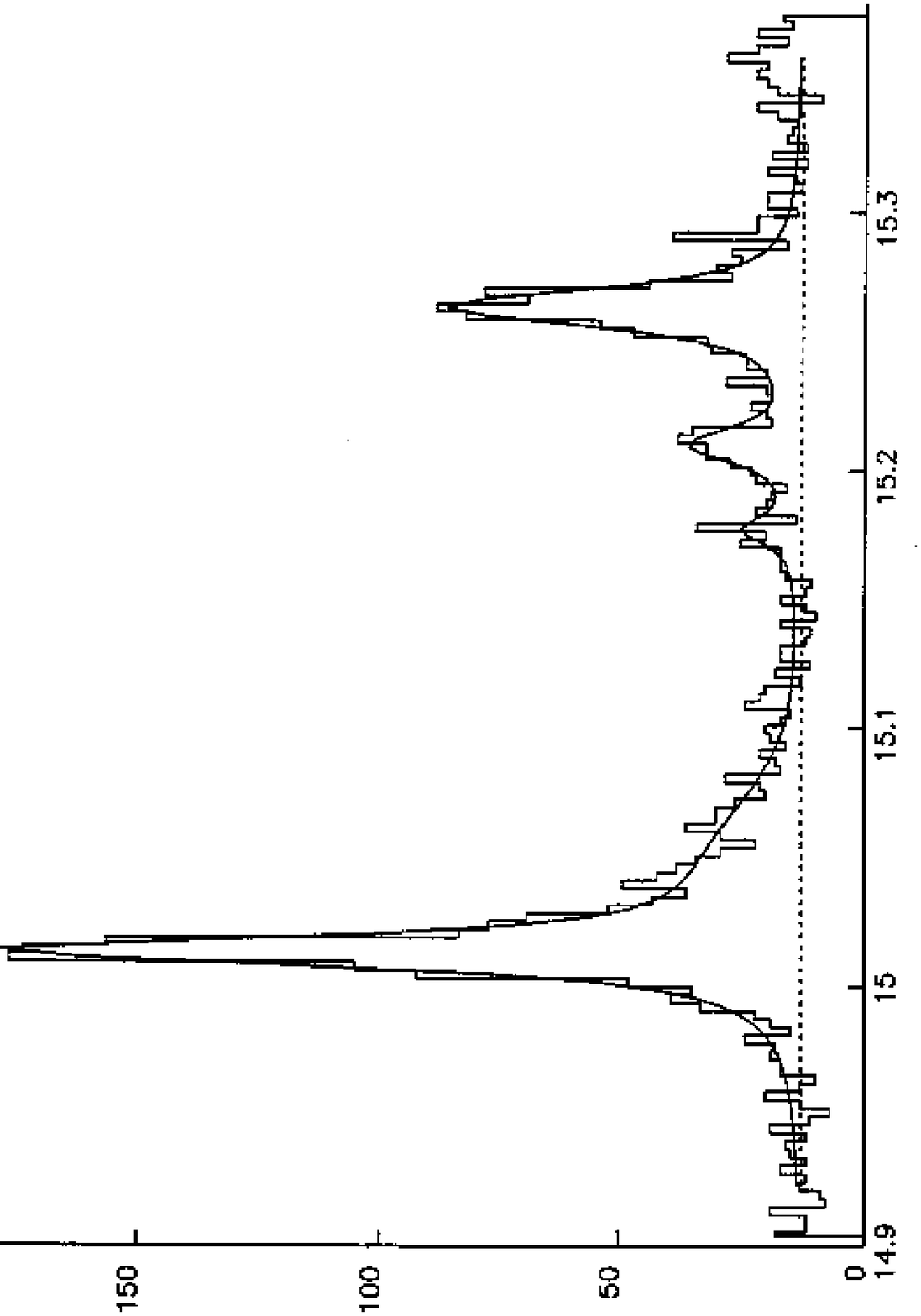}
\caption{ A portion of the FCS spectrum of Active Region 4787/90 at solar coordinates S518, E028 observed on 1987 April 13 at 01:08 UT. The histogram depicts the actual spectral data and the solid curve shows the fit to the four lines: \ion{Fe}{17} $\lambda$15.01, \ion{O}{8} $\lambda$15.18, \ion{Fe}{16} $\lambda$15.21, and \ion{Fe}{17} $\lambda$15.26. }
\end{figure}

\clearpage
\begin{figure}
\includegraphics[width=3.0in]{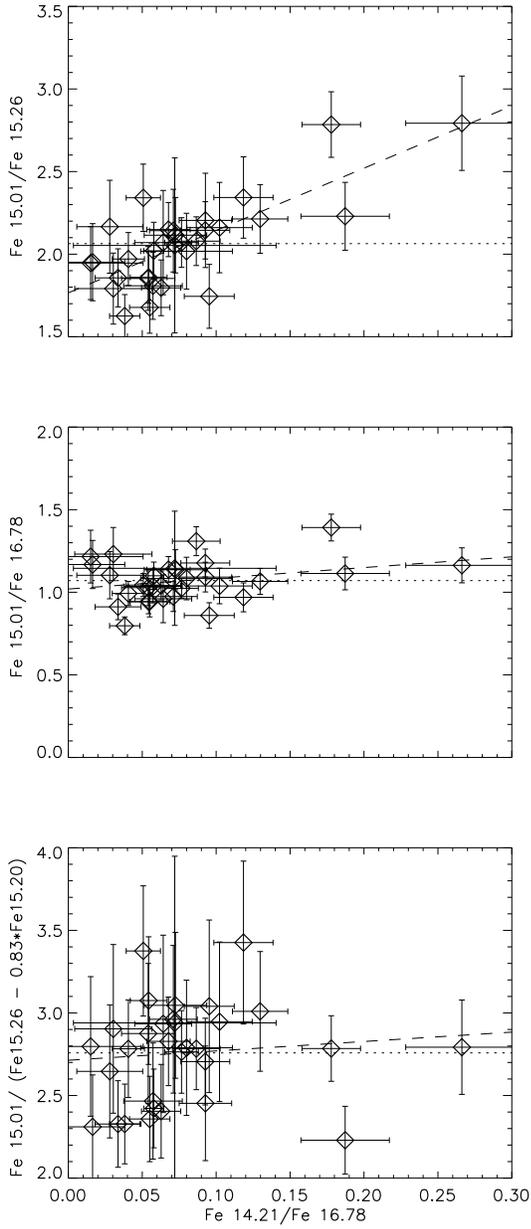}
\caption{Ratios plotted as a function of the temperature indicator (Fe
  XVIII $\lambda$14.21  to \ion{Fe}{17} $\lambda$16.78). The dotted line shows the best flat fit 
and the dashed line shows the best linear fit; ({\it upper}) 3C/3D ratio 
with flat ($ y = 2.06$) and linear ($y = 3.77 x + 1.77$) fits; 
({\it middle}) $\lambda$15.01-to-$\lambda$16.78 ratio with flat ($ y = 1.07$) and linear 
($y = 0.65 x + 1.02$) fits; ({\it lower}) same as ({\it upper}) except that 0.83 times 
the  \ion{Fe}{16} $\lambda$15.21 flux has been subtracted from $\lambda$15.26~\AA\ flux 
for all but the three hottest spectra with flat ($ y = 2.76$) and linear 
($y = 0.57 x + 2.71$) fits.}
\end{figure}

\clearpage
\begin{figure}
\epsscale{.9}
\plotone{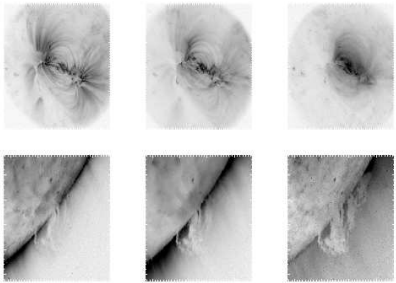}
\caption{\TRACE\ images of active region loops (top) from 1998 May 19 and a 
prominence (bottom) from 1998 May 20 as seen in the 171-\AA, 195-\AA, and 
284-\AA passbands. Note that the loops are fuzzy in the 284-\AA\ image, but 
the absorption edges are sharp. The absorption itself is more pronounced in the 
284-\AA\ image because of the wavelength dependence (Eq. 1).}
\end{figure}

\end{document}